\documentstyle[11pt,newpasp,twoside,epsf]{article}
\def\edcomment#1{\iffalse\marginpar{\raggedright\sl#1\/}\else\relax\fi}
\reversemarginpar

\begin{document}
\title{A solution to a 100-yr problem -- a connection between the 
transition phase in classical novae and intermediate polars} 

\author{A.~Retter}
\affil{School of Physics, University of Sydney, 2006, Australia;
retter@Physics.usyd.edu.au; and Dept. of Physics, Keele University, 
Keele, Staffordshire, ST5 5BG}

\begin{abstract}
One hundred years after the eruption of GK Per and the detection of 
oscillations in its light curve it is still unclear why certain novae 
show a smooth decline in their optical light curves after outburst while 
others have oscillations / minimum during the `transition phase'. Two 
years ago we pointed out a possible connection between this phase in 
classical novae and intermediate polars (IPs), and predicted that novae 
that have the transition phase should be IPs. Chandra observations on 
two recent novae support this prediction. Nova V1494 Aql 99/2, which had 
oscillations during the transition phase, can be classified as an IP, 
while X-ray observations on Nova V382 Vel 1999, whose optical decline 
from outburst was smooth, failed to detect any short-term periodicity. 
About 10\% of the CV population are IPs. This figure is consistent with the 
rarity of the transition phase in novae. We also present a model for this 
idea: the nova outburst disrupts the accretion disc only in IPs because 
their discs are less massive than in non-magnetic systems and their inner 
parts are depleted; the recovery of the disc, a few weeks-months after 
the eruption, causes the formation of dust.

\end{abstract}

\section{Introduction}

The optical light curve of a classical nova is typically charactersed by 
a smooth decline. Certain novae show, however, a DQ Her-like deep minimum 
in the light curve while others have slow oscillations, during the so 
called `transition phase'. The minimum is understood by the presence of 
a dust envelope around the binary system, and it was suggested that the 
oscillations may be connected with the formation of dust blobs moving in 
and out of the line of sight to the nova. It is still unknown, however, 
why only a small fruction ($\sim$15\% of the nova population share this 
peculiar behaviour. 

Retter, Liller \& Gerradd (2000a) suggested a solution for this problem. 
Their observations of Nova LZ Mus 1998, which had oscillations during the 
transition phase, also revealed a few periodicities in its optical light 
curve. LZ Mus was thus classified as an IP candidate. Retter et al. further 
proposed a possible connection between the transition phase and IPs, and 
predicted that novae having a transition phase should be IPs. Recent 
observations of two young novae seem to support this idea.

\section{Observations and Analysis}

The optical light curve of Nova V1494 Aql 1999/2 showed slow oscillations 
with a quasi-period of 7 days (Kiss \& Thomson 2000). A period of 3.2 h 
(presumably the orbital period) was discovered in its optical light curve 
(Retter et al. 2000b). Drake et al. (2001, in preparation) detected a 
2523-s periodicity in two X-ray runs using Chandra. The short period can 
be interpreted as the spin period of the binary system, and the nova is 
very likely an IP.

On the other hand, Nova V382 Vel 1999 had a smooth decline in the optical,
and X-ray observations did not reveal any short-term periodicity. So it is
unlikely that it is an IP.

About 10\% of the cataclysmic variable population are IPs, which is 
consistent with the rarity of the transition phase in novae.

\section{Discussion}

There seems to be a strong connection between the transition phase in 
classical novae and IPs. Two famous examples are Nova GK Per 1901 
(oscillations) and Nova DQ Her 1936 (minimum), which are well-known IPs. 
We suggest that this link is connected with the accretion disc. In IPs 
its inner part is truncated, thus it is less massive than in non-magnetic 
systems. The nova outburst can, therefore, disrupt the disc. Its 
re-establishment is a violent process that forms strong winds that block 
the radiation from the hot white dwarf, and the dust is not destroyed. 
The disc oscillates until finally reaching stability at the end of the 
transition phase. In non-magnetic systems the disc is barely disturbed 
and becomes stable much faster, and in polars there is no disc. In both 
these groups there is no transition phase at all since dust cannot be 
formed.

If the winds are very strong, as might be the case in DQ Her (perhaps since 
its spin period is very short) this leads to a dust minimum. IPs with longer
spin periods should have oscillations.


\begin{references}

\reference Kiss, L.L., \& Thomson, J.R. 2000, A\&A, 355, L9
\reference Retter, A., Liller, W., \& Gerradd, G. 2000a, in "Cataclysmic 
Variables: a 60th Birthday Symposium in Honour of Brian Warner", eds. 
Charles, P., King, A., O'Donoghue, D., Elsevier Science, New Astronomy 
Review, 40/1-2, p65.
\reference Retter, A., Cook, L., Nov\'{a}k, R., Saxton, J.M., Jensen, L.T., 
Kor\v{c}akov\'{a}, D., \& Jan\'{i}k, J. 2000b, IAUC, 7537.

\end{references}
\end{document}